\documentclass[preprint,12pt]{elsarticle}

\usepackage{booktabs}

\usepackage{pifont}
\newcommand{\cmark}{\ding{51}}
\newcommand{\xmark}{\ding{55}}
\newcommand{\tablespace}{\hspace{0.02\textwidth}}

\journal{Computers and Security}

\begin{document}

\begin{frontmatter}
\date{}


\title{SoK: Realistic Adversarial Attacks and Defenses for Intelligent Network Intrusion Detection}


\author[gecad]{Jo{\~{a}}o Vitorino\corref{c1}}
\ead{jpmvo@isep.ipp.pt}

\author[gecad]{Isabel Pra{\c{c}}a\corref{c1}}
\ead{icp@isep.ipp.pt}

\author[gecad]{Eva Maia}
\ead{egm@isep.ipp.pt}

\cortext[c1]{Corresponding author}

\address[gecad]{{Research Group on Intelligent Engineering and Computing for Advanced Innovation and Development (GECAD), School of Engineering, Polytechnic of Porto (ISEP/IPP)}, {4249-015}, {Porto}, {Portugal}}


\begin{abstract}
Machine Learning (ML) can be incredibly valuable to automate anomaly detection and cyber-attack classification, improving the way that Network Intrusion Detection (NID) is performed. However, despite the benefits of ML models, they are highly susceptible to adversarial cyber-attack examples specifically crafted to exploit them. A wide range of adversarial attacks have been created and researchers have worked on various defense strategies to safeguard ML models, but most were not intended for the specific constraints of a communication network and its communication protocols, so they may lead to unrealistic examples in the NID domain. This Systematization of Knowledge (SoK) consolidates and summarizes the state-of-the-art adversarial learning approaches that can generate realistic examples and could be used in real ML development and deployment scenarios with real network traffic flows. This SoK also describes the open challenges regarding the use of adversarial ML in the NID domain, defines the fundamental properties that are required for an adversarial example to be realistic, and provides guidelines for researchers to ensure that their future experiments are adequate for a real communication network.
\end{abstract}



\begin{keyword}
realistic adversarial examples \sep adversarial robustness \sep cybersecurity \sep intrusion detection \sep machine learning
\end{keyword}

\end{frontmatter}


\section{Introduction}
\label{sec:intro}

Modern organizations can benefit from the digital transformation to re-engineer their business processes, integrating control and information systems and automating decision-making procedures. Nonetheless, as organizations become more and more dependent on digital systems, the threat posed by a cyber-attack skyrockets \cite{EuropeanUnionAgencyforCybersecurity2022b}. Every novel technology adds hidden vulnerabilities that can be exploited in multiple attack vectors to disrupt the normal operation of a system. This is particularly concerning for organizations that deal with confidential information and sensitive personal data, or manage critical infrastructure, such as the healthcare and energy sectors \cite{Verizon2022}.

The disruptions caused by a successful cyber-attack can be extremely costly for an organization. In 2022, the average cost of a data breach was reported to be 4.35 million US dollars, an increase of 12.7\% since 2020 \cite{IBMSecurity2022}. This continued growth of both the number of successful cyber-attacks and their associated costs in various sectors and industries denotes that modern organizations face tremendous security challenges. Furthermore, since monitoring a system to detect suspicious activity is not a trivial process and small enterprises commonly fall short of security best practices, most go out of business within 6 months of a breach \cite{Mansfield-Devine2022}.

With financial security and business continuity on the line, it is essential for organizations to improve the way they perform Network Intrusion Detection (NID). This is where Artificial Intelligence (AI), and more specifically Machine Learning (ML), can be incredibly valuable \cite{EuropeanUnionAgencyforCybersecurity2022a}. ML models can originate from numerous algorithms, including tree-based algorithms and deep learning algorithms based on Artificial Neural Networks (ANNs), and can be trained to automate several tasks, from the recognition of patterns and anomalies in network traffic flows to the classification of complex cyber-attacks. The adoption of intelligent cybersecurity solutions can improve resilience and shorten the time required to detect and contain an intrusion by up to 76 days, leading to cost savings of up to 3 million US dollars \cite{IBMSecurity2022}.

However, despite the benefits of ML to tackle the growing number and increasing sophistication of cyber-attacks, it is highly susceptible to adversarial examples: cyber-attack variations specifically crafted to exploit ML models \cite{Szegedy2014}. Even though the malicious purpose of a cyber-attack causes it to have distinct characteristics that could be recognized in a thorough analysis by security practitioners, an attacker can generate specific data perturbations in a network traffic flow to evade detection from intelligent security systems. ML engineers and security practitioners still lack the knowledge and tools to prevent such disruptions, so adversarial attacks pose a major threat to ML and to the systems that rely on it \cite{EuropeanUnionAgencyforCybersecurity2020,SivaKumar2020}.

In recent years, a wide range of adversarial attacks have been developed and researchers have worked on various defenses to protect ML models, but most were not intended for the specific requirements of a communication network and the utilized communication protocols, so they may lead to unrealistic data perturbations in the NID domain. Even though several reviews have been published with comparisons of the strengths and weaknesses of multiple methods \cite{Martins2020,He2023}, they do not address a key aspect: whether or not they could be used in a real communication network. Therefore, there is a lack in the current literature of a systematization of the approaches capable of generating realistic adversarial examples in the NID domain.

This Systematization of Knowledge (SoK) consolidates and summarizes the state-of-the-art adversarial learning approaches that could be applied in real ML development and deployment scenarios with real network traffic flows, and provides guidelines for future research to better address realism. The main Research Question (RQ) to be investigated was:

\begin{itemize}
\item How can adversarial cyber-attack examples be realistically used to attack and defend the ML models utilized in NID?
\end{itemize}

To provide more specific directions for the research, the main RQ was divided into three narrower sub-questions:

\begin{description}
\item[RQ1:] What are the main perturbation crafting processes?
\item[RQ2:] What are the most realistic attack methods?
\item[RQ3:] What are the most reliable defense strategies?
\end{description}

By consolidating the main constraints and limitations of adversarial ML in the NID domain, this SoK intends to guide ML engineers and security practitioners to improve their methods and strategies according to the constraints of their specific communication networks. For that purpose, it is organized into multiple sections. Section \ref{sec:methdlgy} describes the adopted research methodology. Sections \ref{sec:perturb}, \ref{sec:attack}, and \ref{sec:defense}, summarize and discuss the findings of RQ1, RQ2, and RQ3, respectively. Section \ref{sec:future} describes the open challenges regarding the main RQ and provides guidelines for future research. Finally, Section \ref{sec:conclsn} presents the concluding remarks.


\section{Research Methodology}
\label{sec:methdlgy}

The research was based on the Preferred Reporting Items for Systematic Reviews and Meta-Analyses (PRISMA) \cite{Moher2015}, which is a standard reporting guideline that aims to improve the transparency of literature reviews. Search terms were used in reputable bibliographic databases, and several inclusion and exclusion criteria were defined to screen the found publications. Since screening the titles and abstracts of the publications was sufficient to assess their eligibility, their full texts were directly reviewed without further exclusion rounds being necessary.

After a careful initial analysis of the literature, several search terms were chosen to address the formulated RQs. To ensure a comprehensive coverage of relevant publications within the scope of adversarial ML, the \textit{adversarial} keyword was combined with other suitable terms like \textit{perturbation}, \textit{attack}, and \textit{defense}. Concepts closely related to NID, such as \textit{anomaly detection}, \textit{cyber-attack classification}, \textit{wireless} and \textit{IoT} networks, were also considered. Table \ref{tab:rmterms} provides an overview of the defined search terms.

\renewcommand{\arraystretch}{1.2}
\begin{table}
\caption{Defined search terms.}
\label{tab:rmterms}
\begin{tabular}{p{0.33\textwidth} p{0.60\textwidth}}
\toprule
Scope & Terms \\
\midrule
Adversarial & \textit{adversarial} \\
Learning & (\textit{learning} OR \textit{example} OR \textit{perturbation} OR \textit{attack} OR \textit{defense}) \\
Network & (\textit{network} OR \textit{wireless} OR \textit{IoT}) \\
Intrusion & (\textit{intrusion} OR \textit{anomaly} OR \textit{cyber-attack}) \\
Detection & (\textit{detection} OR \textit{classification}) \\
\bottomrule
\end{tabular}
\end{table}

The primary search source was Science Direct \cite{SourceElsevier}, which is a large bibliographic database of scientific journals and conference proceedings provided by the internationally recognized publisher Elsevier. Due to their relevance for scientific literature of ML, computing, software engineering, and information technology, the search also included the digital libraries of the Association for Computing Machinery (ACM) \cite{SourceACM}, the Institute of Electrical and Electronics Engineers (IEEE) \cite{SourceIEEE}, and the Multidisciplinary Digital Publishing Institute (MDPI) \cite{SourceMDPI}. It is important to note that the PRISMA backward snowballing process of checking the references of the findings led to additional records that were not directly obtained from these databases.

Since adversarial ML is an active research field, the search was limited to peer-reviewed publications in journals or conference proceedings from 2017 onwards. It included recent works addressing the use of adversarial ML in the NID domain, as well as surveys and reviews that addressed key developments, which led to additional publications. The findings that were duplicated in multiple databases were removed, and those that were not directly applied to the NID domain or did not introduce a novel method or strategy were excluded. Table \ref{tab:rmcrits} provides an overview of the inclusion and exclusion criteria that were defined to screen the found publications.

\renewcommand{\arraystretch}{1.2}
\begin{table}
\caption{Defined inclusion and exclusion criteria.}
\label{tab:rmcrits}
\begin{tabular}{p{0.53\textwidth} p{0.40\textwidth}}
\toprule
Inclusion Criteria & Exclusion Criteria \\
\midrule
IC1: Peer-reviewed journal article or conference paper & EC1: Duplicated publication \\
IC2: Available in the English language & EC2: Not applied to NID \\
IC3: Published from 2017 onwards & EC3: Not a novelty \\
IC4: Addressed adversarial ML for NID & EC4: Full text not available \\
\bottomrule
\end{tabular}
\end{table}

A total of 936 records were initially retrieved by applying the query to the contents of the publications stored in the selected databases. After removing duplicates and performing the screening phase, 139 records were excluded because they mentioned NID but were not directly applied to the NID domain. Furthermore, another 703 records were excluded because they did not present novel approaches. Despite performing experiments with different datasets and different contexts, these records used previously published methods and strategies without relevant modifications. The 703 records correspond to over 75\% of the found publications, which demonstrates that it is difficult for researchers to find innovative approaches in a regular search in these databases, and further highlights the necessity for a systematization of the most relevant advances in this research field.

\begin{figure*}
    \centering
    \includegraphics[scale=0.70]{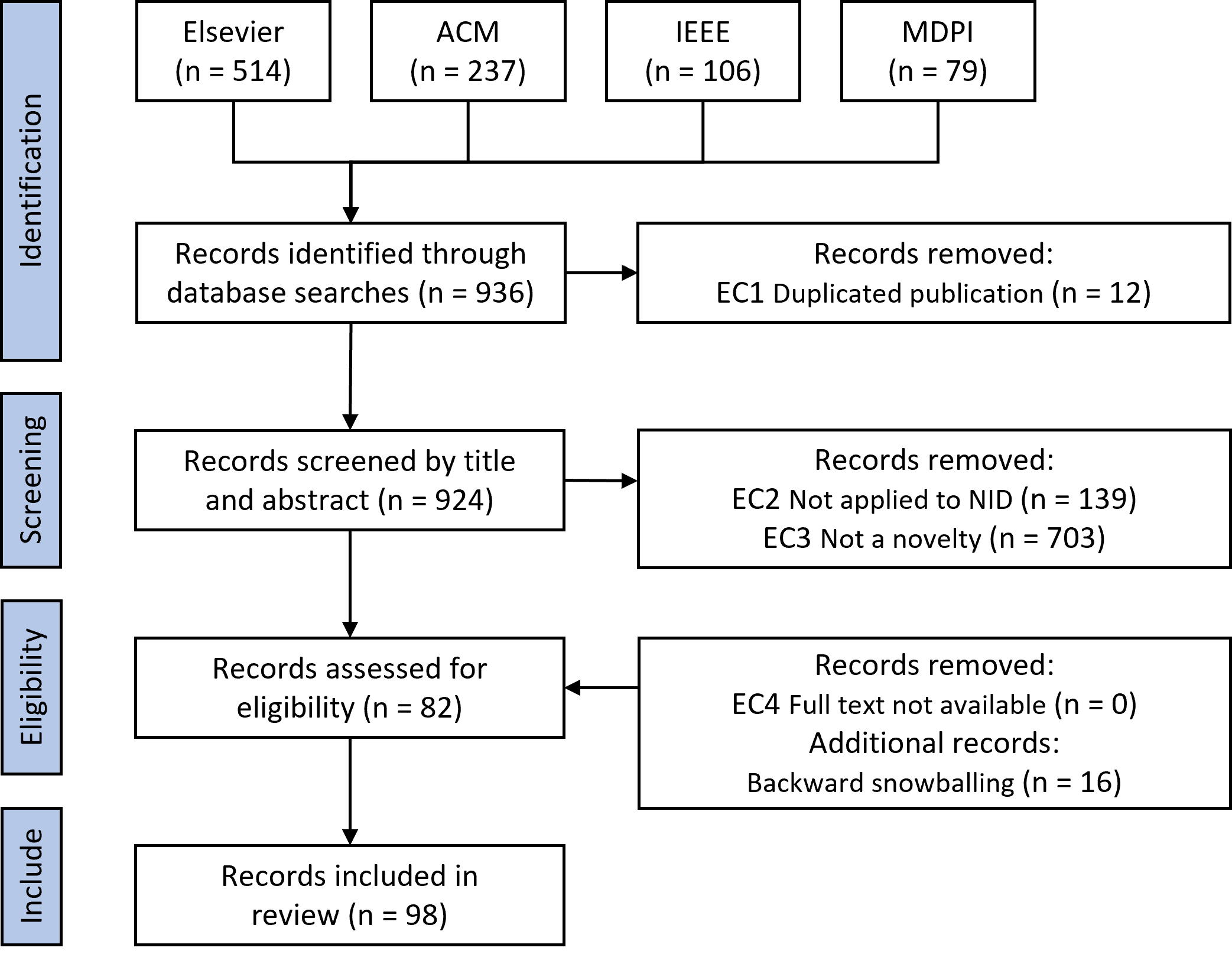}
    \caption{PRISMA search process.}
    \label{fig:prisma}
\end{figure*}

The remaining 82 records, which correspond to only 9\% of the found publications, provided relevant aspects for the use of adversarial ML in the NID domain. Their content and references were checked and 16 additional publications were found by performing backward snowballing. Therefore, 98 publications were included in the review (see Figure \ref{fig:prisma}). The publications were independently reviewed by each author of this SoK to extract the key developments and key takeaways for NID, and then their notes were consolidated and systematized to create this manuscript.

\section{Data Perturbations}
\label{sec:perturb}

ML has been increasingly used to make digital systems more intelligent, but it is not flawless. For instance, if an ML model is trained with non-representative data that has missing or biased information, it may become underfit, performing poorly on both its training data and new data, or even overfit, performing very well on its training data but still poorly on previously unseen testing data \cite{Liu2019}. These generalization errors can be quickly noticed during the development of an intelligent system with ML models, and better results can be achieved by improving data quality and fine-tuning the models \cite{Alaoui2022}. However, even if a model generalizes well to the testing data, it is not guaranteed to always have a stable performance. During the inference phase, when it is deployed to make predictions on live data, it may sometimes behave unexpectedly with seemingly ordinary data samples \cite{Salman2020,Thakkar2020}.

In a set of very similar samples of the same class, a model may correctly classify all but one. That specific sample may be assigned to a completely different class with a high confidence score because the model wrongly considers that it is different from the others. Ultimately, this unexpected behavior is caused by some unnoticed generalization errors during the model’s training phase \cite{Stutz2019}. Since a training set does not cover all the samples that a model will encounter in its inference phase when deployed in a real system, the model will inevitably learn some simplifications of the decision boundaries that lead to incorrections in its internal reasoning \cite{Fawzi2015,Sabour2016}. These incorrections can be hard to notice because the intricate mechanics of ML models cause the misclassifications to only occur in very specific samples, which are designated as \textbf{adversarial examples} \cite{Szegedy2014}.

An adversarial example may have very subtle perturbations that are almost imperceptible to humans but make it significantly different from regular samples to an ML model. Such adversarial perturbations can occur naturally in faulty data recordings with incorrect readings, but they can also be specifically crafted with specialized inputs to exploit the generalization errors \cite{Goodfellow2015,Kurakin2017}. Even though all ML models are inherently susceptible to adversarial examples, different models will learn distinct simplifications of the target domain and create distinct decision boundaries. Hence, some models may be more vulnerable to perturbations in a certain feature than others, presenting model-specific edge cases that are hard to detect and address \cite{Papernot2016b,Tabacof2016}.

Due to the advances in computer vision technologies and their increasing use across various industries, the major developments in adversarial ML have been focused on the image classification domain and are then adapted to other domains \cite{Ren2020,Zhang2020a}. In adversarial images, the perturbed features are pixels with a value freely assigned from 0 to 255, but it is pertinent to understand how these research efforts can be applied in cybersecurity solutions and if the concepts are transferable to a NID system in a real communication network. In the current literature, the perturbations that turn a regular sample into an adversarial example can be crafted using two main concepts: an \textbf{adversarial patch} that heavily modifies a few features, and an \textbf{adversarial mask} that slightly modifies many or all features \cite{Wiyatno2019}.

\begin{figure*}
    \centering
    \includegraphics[scale=0.9]{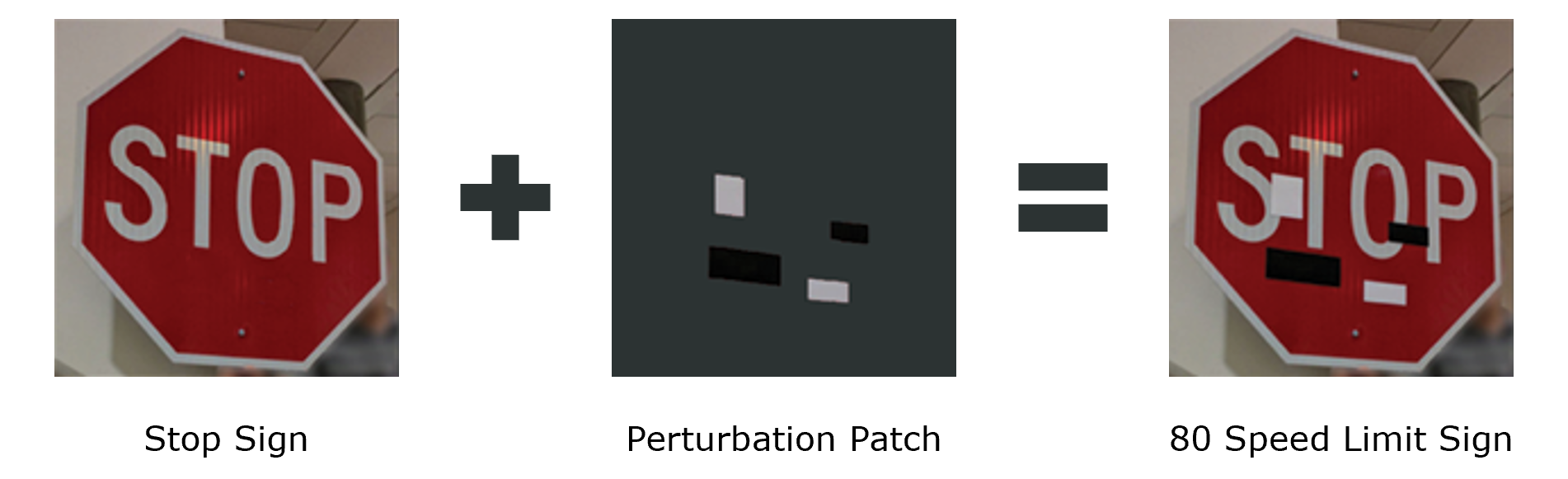}
    \caption{Adversarial perturbation via a patch, based on \cite{Eykholt2018}.}
    \label{fig:advpatch}
\end{figure*}

Adversarial patches are the most straightforward way to disrupt a cyber-physical system. Since live data from a physical environment is not easily controllable, there is a greater risk for an ML model to be affected by faulty input data, either naturally occurring or purposely created \cite{Li2020}. For instance, for a model trained to classify street signs, a perturbed sample of a stop sign with small black and white patches can be misclassified as a completely unrelated sign (see Figure \ref{fig:advpatch}) \cite{Eykholt2018}. These patches are devised to cause the model to make a mistake when it encounters the sign at a certain angle, although a human would still recognize a stop sign \cite{Brown2017}.

\begin{figure*}
    \centering
    \includegraphics[scale=0.9]{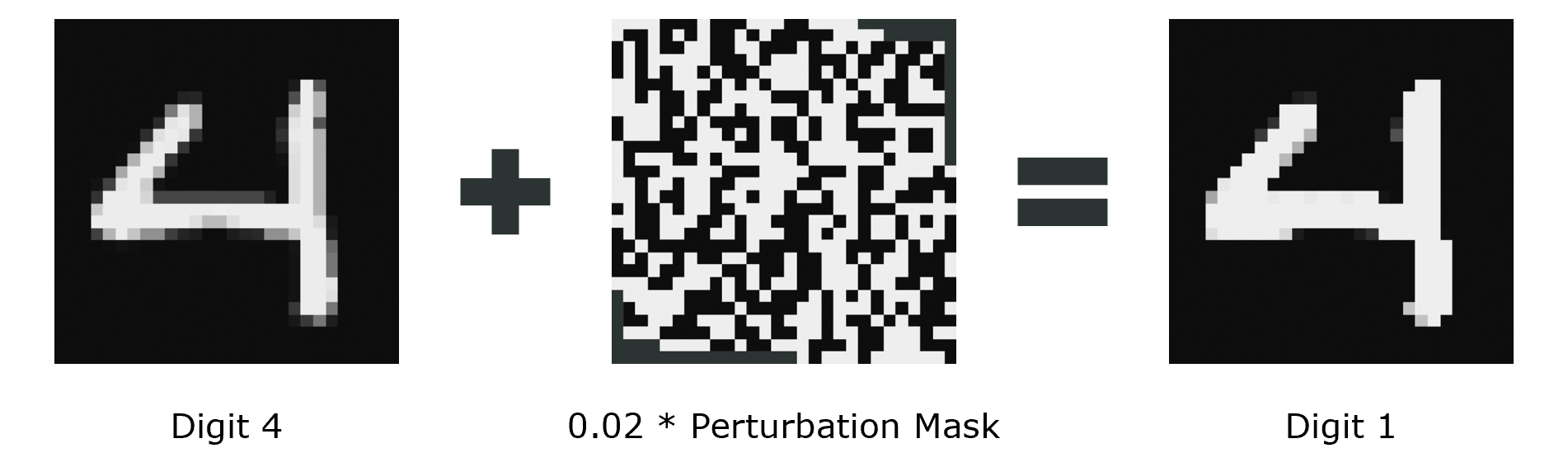}
    \caption{Adversarial perturbation via a mask, based on \cite{Edwards2020}.}
    \label{fig:advmask}
\end{figure*}

Despite being harder to apply adversarial masks in physical environments, they are very well-suited for digital systems. For instance, for a model that performs handwritten digit recognition, a picture of a digit with a subtle change to several pixels can be misclassified as another digit (see Figure \ref{fig:advmask}) \cite{Edwards2020}. Such model can have a wide range of applications, from certified documents and bank check processing to authentication via a picture of an identification document. If a person applies a filter that has a built-in adversarial mask before submitting the requested document, the automated verification systems that rely on this model can be deceived \cite{Rosenberg2021}. Furthermore, there are even some adversarial masks that exploit the intrinsic vulnerabilities of ML models and turn every image of well-established datasets into an adversarial image, which denotes that adversarial examples might not be as difficult to create as previously thought \cite{Moosavi-Dezfooli2017}.

\begin{figure*}
    \centering
    \includegraphics[scale=0.78]{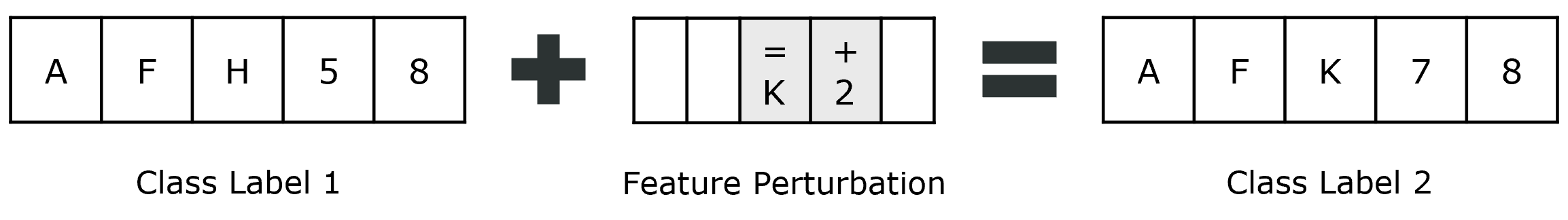}
    \caption{Adversarial perturbation on tabular data, based on \cite{Merzouk2022}.}
    \label{fig:advtab}
\end{figure*}

Even though most developments in the adversarial ML area of research have addressed image classification, the susceptibility of ML models to these examples has also been noticed in other domains with different data types, such as audio, text, tabular data, and time series \cite{Papernot2016b,Yuan2019}. For the NID domain, adversarial perturbations must follow a tabular format, where each feature is a categorical or numerical variable representing a characteristic of network traffic \cite{Papadopoulos2021,Hashemi2019}. This tabular data format requires more complex perturbations, but they can also be based on the concepts utilized for images. For a tabular classification model, a patch-like perturbation could fully replace the values of categorical variables, which may include the communication protocol or the endpoint port number, and a mask-like perturbation could slightly increase or decrease the values of numerical variables, such as the amount of sent packets or the download to upload ratio (see Figure \ref{fig:advtab}) \cite{Merzouk2022,Peng2019}. Nonetheless, not all perturbations are suitable for the NID domain because there are specific constraints that must be complied with.

In contrast with the pixels of an image, each tabular feature may have a different range of possible values, according to the characteristic it represents. Furthermore, a feature may also be highly correlated to several others, being required to exhibit specific values depending on the other characteristics of a sample \cite{Zhou2022}. For instance, a Slowloris is a Denial-of-Service (DoS) attack that attempts to overwhelm a web server by opening multiple connections and maintaining them as long as possible. A flow utilized in this cyber-attack must use the Transmission Control Protocol (TCP) and the Push (PSH) flag to keep the connection open on the port number 80, its endpoint \cite{Shorey2018}.

A very relevant characteristic of this flow is its packet Inter-Arrival Time (IAT), which represents the elapsed time between the arrival of two subsequent packets and may be represented as two features: the minimum and maximum IAT. The flow may have a varying IAT between 20 and 30 seconds to appear as arbitrary benign traffic instead of scheduled packets just to keep the connection open. In a certain network, a longer IAT cannot be used because the web server being attacked is configured with a timeout to close connections after 30 seconds of inactivity, which is a very common web application security measure \cite{Al-Qudah2010}.

However, throughout the literature, various studies provide adversarial cyber-attack examples crafted via patch-like and mask-like perturbations as direct input to an ML model without questioning if they are viable for a real communication network \cite{Martins2020,Vitorino2023a}. This may result in misleading evaluations where the ML models are tested against unrealistic examples that they will not encounter in a real deployment scenario with real network traffic. 

Due to their lack of constraints, it is very difficult to transfer the perturbation crafting processes of the image classification domain to the NID domain. A patch-like crafting process could be performed, changing the flow from a TCP connection to another protocol, or from port number 80 to another port, but these modifications would not be useful for a lengthy DoS. The communication protocol, the connection flag, and the port must remain the same, otherwise the crafted example will no longer be a flow of the Slowloris class \cite{Chauhan2021,Lin2022}. Likewise, a mask-like crafting process may increase or decrease the values of the minimum and maximum IAT, but not all perturbations will be suitable for a real communication network.

\begin{figure*}
    \centering
    \includegraphics[scale=0.78]{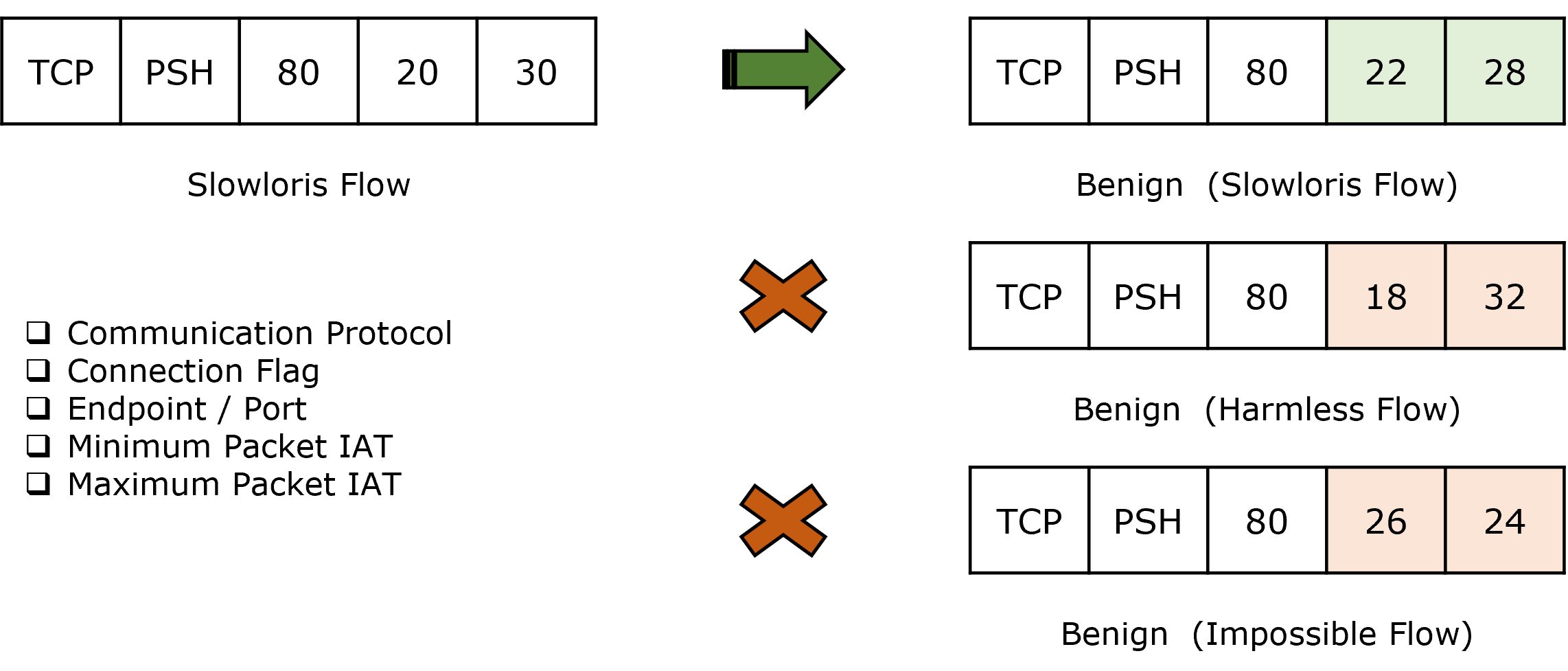}
    \caption{Adversarial perturbation on a network traffic flow, based on \cite{Vitorino2023}.}
    \label{fig:advflow}
\end{figure*}

Three diferent examples may be generated for the considered Slowloris flow: the first with a minimum IAT of 22 and a maximum IAT of 28 seconds, the second with 18 and 32, and the third with 26 and 24. Even though all three may deceive an ML model and be misclassified as belonging to the benign class, only the first is a harmful Slowloris flow. The second example is actually a harmless flow because the considered web server will terminate the connection at the 31st second, before a packet is received at the 32nd second, preventing the functionality of this type of DoS \cite{Vitorino2022}. In turn, the third example would not even be possible in a real communication network because a flow with packets at least every 26 seconds cannot also have packets at most every 24 seconds (see Figure \ref{fig:advflow}).

Despite all examples following similar mask-like perturbations of increasing and decreasing some numerical variables of the flow, they would lead to very different outcomes in a communication network and only one example could be used against a real NID system. Therefore, a successful adversarial attack is not guaranteed to be a successful cyber-attack \cite{Vitorino2023}.

To ensure that an adversarial example represents a real network traffic flow that can be transmitted through a real communication network, the constraints of the utilized communication protocols and the malicious purpose and functionality of a cyber-attack must be taken into account when generating the perturbations \cite{Apruzzese2021,McCarthy2021}. Nonetheless, despite the current difficulty in creating realistic adversarial cyber-attack examples, the growing popularity of adversarial ML is leading to the development of novel methods to attack various types of algorithms, which is very concerning for the security of intelligent systems \cite{Martins2020,Papadopoulos2021,SivaKumar2020}.


\section{Attack Methods}
\label{sec:attack}

The susceptibility of ML to adversarial examples can be exploited for diverse malicious purposes with methods that automatically generate the required perturbations. An attacker targeting an intelligent system may use multiple methods to perform a wide range of attacks, which can be divided into two primary categories: \textbf{poisoning attacks} during a model’s training phase, and \textbf{evasion attacks} during the inference phase \cite{Pitropakis2019}.

Poisoning attacks inject adversarial examples in a model’s training data to compromise its internal reasoning and decision boundaries. These attacks can perform \textbf{model corruption} that make it completely unusable, or even introduce \textbf{hidden backdoors} that make it exhibit a biased behavior in specific samples, which is difficult to detect and explain because the model only deviates from its expected behavior when triggered by very specific perturbations \cite{Gu2017,Li2022}. This is a serious security risk for organizations that heavily rely on third-party datasets or outsource their cybersecurity solutions, such as the development of facial recognition models for biometric authentication systems \cite{Chen2017,Wang2019}. Nonetheless, since NID systems are commonly developed in secure environments with thoroughly verified network traffic data, an external attacker does not usually have access to an ML model to compromise it during its training phase \cite{Apruzzese2021,He2023}.

On the other hand, evasion attacks use adversarial examples to deceive a vulnerable model after it has been deployed. The misclassifications caused by these attacks can be directly used to \textbf{evade detection} from an intelligent security system, or for more complex goals, such as \textbf{membership inference} and \textbf{attribute inference} to check if a model was trained with a certain sample and certain features, \textbf{model inversion} to reconstruct a training set, and \textbf{model extraction} to steal its functionality and replicate it in a substitute model \cite{Fredrikson2015,Qiu2019,Shokri2017}. If confidential or proprietary information is used to train a model, an attacker can cause significant damage to an organization by gathering that information during the inference phase \cite{Hitaj2017,Veale2018}. Even though a model must be queried many times to obtain the information, advances in wireless and IoT technologies are making NID systems process larger and larger amounts of network traffic, which substantially increases query opportunities and therefore the feasibility of evasion attacks \cite{Aiken2019,Flowers2020,Ibitoye2019}.

In recent years, numerous methods have been created to automate the misclassification attempts for evasion attacks. A method may require access to a model in one of three possible settings: \textbf{black-box}, \textbf{gray-box}, and \textbf{white-box}. The first is model-agnostic and solely queries a model’s predictions, whereas the second may also require knowledge of its architecture or the utilized features, and the third needs full access to its internal parameters \cite{Chakraborty2021,Li2020}. Additionally, a black-box or gray-box method may solely use class predictions, a \textbf{decision-based} approach, or require a model to output the confidence scores of the predictions, a \textbf{score-based} approach \cite{Ilyas2018,He2023}. These characteristics affect the choice of an adversarial method because it must be able to attack the targeted model and system, while also being useful to the fulfillment of the goals of the attacker.

Since the focus of adversarial ML has been image classification, the common attack approach is to freely exploit the internal gradients of an ANN in a white-box setting \cite{Pitropakis2019,Yuan2019}. Consequently, most state-of-the-art methods do not support other settings nor other models, which severely limits their applicability to other domains. Considering that a deployed NID system is securely isolated, having full access to a model and its feedback is highly unlikely, and an attacker will only know if a certain example evades detection if the entire cyber-attack is successfully completed. This interaction corresponds to a decision-based approach in black-box or gray-box settings, depending on the available system information about the utilized model and feature set \cite{Apruzzese2021,Rosenberg2021}. Furthermore, various other types of ML models can be used for classification tasks with tabular data. For instance, tree-based algorithms and ensembles like Random Forest (RF) are remarkably well-established for NID, but are also susceptible to adversarial attacks \cite{Belavagi2016,Liu2019,Primartha2018,Pujari2022}. Therefore, an attacker will have to resort to methods that support these models and all the specificities of NID.

Various adversarial evasion attack methods have been made open-source software and have started being used to target the ML models of intelligent NID systems. Table \ref{tab:attmets} summarizes the characteristics of the most relevant methods of the current literature that have been used in NID, noting if they could potentially fulfill the constraints of complex communication networks. Even though some methods were introduced as suitable for a black-box setting, they require knowledge of the utilized feature set to determine how which feature will be perturbed, so they were categorized as being in the gray-box setting. The ‘Scores’ keyword corresponds to models that can output confidence scores for a score-based approach. In turn, the ‘Gradients’ keyword corresponds to models that provide full access to their internal loss gradients, which includes ANNs.

\renewcommand{\arraystretch}{1.2}
\begin{table*}
\caption{Characteristics of relevant adversarial evasion attack methods.}
\label{tab:attmets}
\begin{tabular}{p{0.22\textwidth} p{0.19\textwidth} p{0.16\textwidth} p{0.15\textwidth} p{0.13\textwidth}}
\toprule
Method & Setting & Models & Constraints & Reference \\
\midrule
BIM & White-box & Gradients & \tablespace\xmark & \tablespace\cite{Kurakin2017} \\
C\&W & White-box & Gradients & \tablespace\xmark & \tablespace\cite{Carlini2017} \\
DeepFool & White-box & Gradients & \tablespace\xmark & \tablespace\cite{Moosavi-Dezfooli2016} \\
FGSM & White-box & Gradients & \tablespace\xmark & \tablespace\cite{Goodfellow2015} \\
Hierarchical & White-box & Gradients & \tablespace\xmark & \tablespace\cite{Zhou2022} \\
Houdini & White-box & Gradients & \tablespace\xmark & \tablespace\cite{Cisse2017} \\
JSMA & White-box & Gradients & \tablespace\cmark & \tablespace\cite{Papernot2016} \\
PGD & White-box & Gradients & \tablespace\xmark & \tablespace\cite{Madry2018} \\
Structured & White-box & Gradients & \tablespace\xmark & \tablespace\cite{Xu2019} \\
GSA-GAN & Gray-box & Scores & \tablespace\xmark & \tablespace\cite{Wang2022} \\
IDS-GAN & Gray-box & Scores & \tablespace\xmark & \tablespace\cite{Lin2022} \\
Polymorphic & Gray-box & Scores & \tablespace\cmark & \tablespace\cite{Chauhan2021} \\
A2PM & Gray-box & Any & \tablespace\cmark & \tablespace\cite{Vitorino2022} \\
DoSBoundary & Gray-box & Any & \tablespace\cmark & \tablespace\cite{Peng2019} \\
BFAM & Black-box & Scores & \tablespace\xmark & \tablespace\cite{Zhang2020} \\
BMI-FGSM & Black-box & Scores & \tablespace\xmark & \tablespace\cite{Lin2020} \\
OnePixel & Black-box & Scores & \tablespace\cmark & \tablespace\cite{Su2019} \\
RL-S2V & Black-box & Scores & \tablespace\xmark & \tablespace\cite{Dai2018} \\
WGAN & Black-box & Scores & \tablespace\xmark & \tablespace\cite{Arjovsky2017} \\
ZOO & Black-box & Scores & \tablespace\xmark & \tablespace\cite{Chen2017a} \\
Boundary & Black-box & Any & \tablespace\xmark & \tablespace\cite{Brendel2018} \\
CGAN & Black-box & Any & \tablespace\xmark & \tablespace\cite{Mirza2014} \\
CVAE & Black-box & Any & \tablespace\xmark & \tablespace\cite{Sohn2015} \\
GADGET & Black-box & Any & \tablespace\xmark & \tablespace\cite{Rosenberg2018} \\
HopSkipJump & Black-box & Any & \tablespace\xmark & \tablespace\cite{Chen2020} \\
Optimization & Black-box & Any & \tablespace\xmark & \tablespace\cite{Cheng2019} \\
\bottomrule
\end{tabular}
\end{table*}

Several methods initially developed for the generation of adversarial images have been adapted to generate adversarial network traffic flows. However, most do not account for the constraints of the utilized communication protocols nor the functionalities of the cyber-attacks, so only a few could potentially generate realistic examples \cite{He2023,Pujari2022}.

Both the Jacobian-based Saliency Map Attack (JSMA) \cite{Papernot2016} and the One Pixel attack \cite{Su2019} were developed to attack image classification models, but their perturbation crafting processes could be used to preserve the structure of a traffic flow. The former minimizes the number of modified pixels, requiring full access to the internal gradients of an ANN in a white-box setting, whereas the latter only modifies a single pixel, based on the confidence scores of a model in a black-box setting. These methods only perturb the most appropriate features in a decision boundary without affecting the remaining ones, which can preserve the correlations between most features of a flow.

Nonetheless, these methods freely generate the perturbations for the few modified features. When adapted to network flows, this lack of constraints could lead to values that are incompatible with the remaining features, which would result in mostly harmless or impossible flows and only a few occasional realistic examples created by chance. To generate high-quality examples on a more regular basis, some methods have been specifically developed to tackle the constraints of the NID domain.

The Polymorphic attack \cite{Chauhan2021} addresses the preservation of original class characteristics to create examples compatible with a cyber-attack’s purpose. A feature selection algorithm is applied in a gray-box setting to obtain the most impactful features for the distinction between benign traffic and cyber-attack classes in a dataset. Then, the remaining features, which are considered non-relevant for the functionality of a cyber-attack, are perturbed by a Wasserstein Generative Adversarial Network (WGAN) \cite{Arjovsky2017}. Despite the WGAN not accounting for the constraints of the remaining features, the most important features of each class are not modified, so the main characteristics required for a successful cyber-attack may be preserved.

The distinction between benign traffic and cyber-attack classes was further explored in the DoSBoundary attack \cite{Peng2019} and the Adaptative Perturbation Pattern Method (A2PM) \cite{Vitorino2022}. Both iteratively optimize the perturbations that are performed on each feature of a traffic flow according to the constraints of a communication network and the functionality of each cyber-attack class. The former requires expert knowledge to manually configure the specific perturbations of each feature, whereas the latter only needs to know the utilized feature set and relies on adaptative patterns to learn the characteristics of each feature. Despite these methods requiring many queries to a model and knowledge of the feature set, which corresponds to a gray-box setting, they can generate constrained adversarial examples that preserve the correlations between the features of a network traffic flow.

Due to the different characteristics of existing methods and diverse goals of attackers, efforts are being made to systematize the possible attack vectors in the Adversarial Threat Landscape for Artificial-Intelligence Systems \cite{WebMITRE} knowledge base, and to complement it with case studies and demonstrations based on real-world observations. As novel adversarial methods continue to be developed, it is becoming essential to raise awareness of the diverse strategies that attackers can use to exploit ML models and the security risks they pose to modern organizations.


\section{Defense Strategies}
\label{sec:defense}

\begin{figure*}
    \centering
    \includegraphics[scale=0.68]{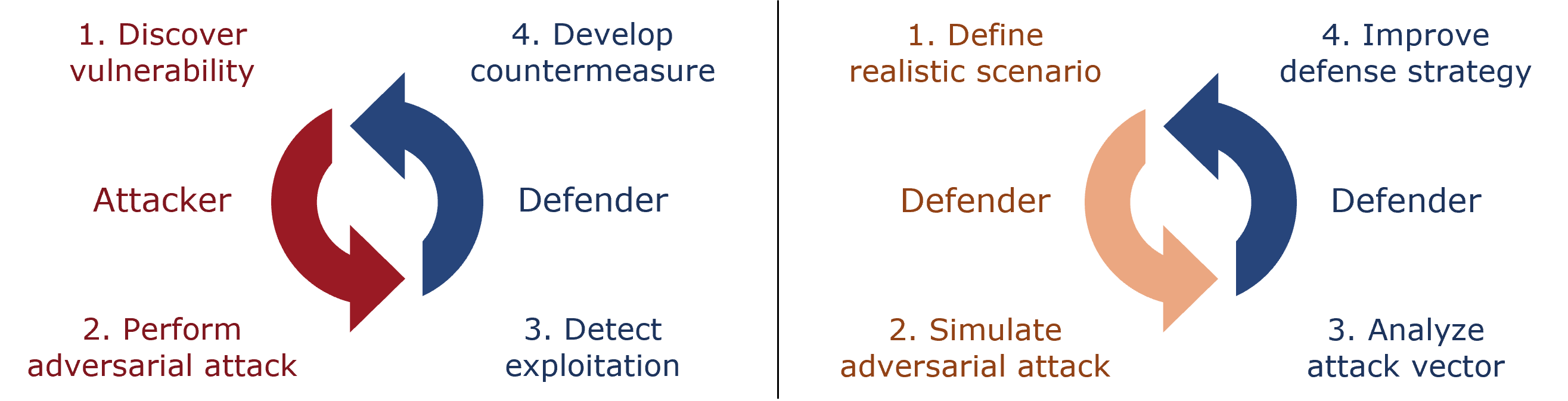}
    \caption{Adversarial arms race (left) and security-by-design (right), based on \cite{Biggio2014}.}
    \label{fig:defloop}
\end{figure*}

The growing ML attack surface led to a never-ending arms race where attackers continuously exploit newly discovered vulnerabilities and defenders develop countermeasures against each novel threat. However, the defenders are always a step behind because it can take a long time until the effects of an attack are detected, and then it is difficult to retrace it and develop a specific countermeasure \cite{EuropeanUnionAgencyforCybersecurity2022}. To get ahead of attackers, organizations should follow a security-by-design development approach and proactively search for vulnerabilities themselves. By simulating adversarial attacks in realistic scenarios and analyzing entire attack vectors, ML engineers and security practitioners can anticipate possible threats and use that knowledge to preemptively revise and improve their defense strategy (see Figure \ref{fig:defloop}) \cite{Biggio2014}.

A defense strategy can combine multiple techniques to address different security concerns. Due to their proven value against several adversarial attacks, some defenses have been standardized across the scientific literature, divided into two primary categories: \textbf{proactive defenses} during a model's training phase, and \textbf{reactive defenses} during the inference phase \cite{Yuan2019}.

Regarding reactive defenses, they attempt to mitigate the effects of corrupted data on a model’s predictions by safely processing its input and output data. These defenses can rely on several preprocessing techniques, such as \textbf{data denoising} and \textbf{feature squeezing} to reduce the search space for an attack, and postprocessing techniques, such as mechanisms that deal with \textbf{model uncertainty} and require predictions with \textbf{high confidence} scores \cite{Chakraborty2021,Qiu2019,Smith2018}. Even though reactive defenses can be valuable against both erroneous data and adversarial attacks purposely exploiting a model, they represent an additional software layer that attempts to encapsulate a vulnerable model. This layer is always needed for a NID system to convert the recorded network traffic into the utilized feature set and then convert the predictions of one or more models into relevant warnings and alerts, but it does not fully protect those models \cite{Apruzzese2021,Thakkar2020}.

On the other hand, proactive defenses tackle the susceptibility of ML to adversarial examples, aiming to reduce the vulnerabilities and intrinsically improve a model’s robustness against adversarial examples during its training phase. These defenses include several techniques, such as \textbf{adversarial training} with perturbed samples, \textbf{regularization} to better calibrate the learning process, and \textbf{defensive distillation} to create smaller models less sensitive to data variations \cite{Goodfellow2015,Martins2020,Papernot2016a,Zhang2020a}. It is not yet clear how to completely resolve this susceptibility and achieve an adversarially robust generalization in a classification task, but progress is being made in robustness research with regularization and optimization techniques \cite{Bai2021,Schmidt2018,Khamis2020}. This gives ML engineers and security practitioners better tools to address ML security during the entire lifecycle of an intelligent system, including its development, testing, deployment, and maintenance phases.

Most proactive defenses are focused on improving the robustness of deep learning algorithms based on ANNs against evasion attacks with adversarial images \cite{Feinman2017,Ganin2017,Miller2020,Tramer2018}, although some also take measures against backdoors \cite{Li2022,Wang2019}. Despite ANN defenses being difficult to apply to other models and domains, the protection of tree-based algorithms has been drawing attention for intelligent cybersecurity solutions \cite{Anthi2021,Apruzzese2020}. Some defenses have been developed to improve the robustness of entire tree ensembles at once \cite{Chen2021,Kantchelian2016}, whereas others address each individual decision tree at a time \cite{Chen2019,Vos2021}. Still, proactive defenses often trade-off some performance on regular samples to improve performance on adversarial examples. This trade-off affects the choice of a defense strategy because the utilized techniques need to balance adversarial robustness and generalization to regular network traffic.

Defense strategies continue to be enhanced with better techniques, but the most effective and widespread defense is still adversarial training because it anticipates the data variations that an ML model may encounter when it is deployed \cite{Andriushchenko2020,Fu2021,Shafahi2019}. Augmenting a training set with examples created by one or more adversarial attack methods enables a model to learn additional characteristics that the samples of each class can exhibit, so it becomes significantly harder to deceive it. This augmented training data with more data variations can improve a model’s robustness not only against attack methods similar to the utilized ones, but also against a wide range of attacks that perform different data perturbations \cite{Bai2021,Shafahi2020,Zhao2022}.

Nonetheless, augmenting a model’s training set with the examples created by adversarial attack methods may not be as beneficial as it seems. Even though it is meant to improve robustness, training with unrealistic samples will make a model learn distorted characteristics that will not be exhibited by regular samples \cite{Stutz2019}. This raises a major security concern because including unrealistic data in a training set can not only be detrimental to a model’s generalization, but also lead to accidental data poisoning and to the introduction of hidden backdoors that leave a model even more vulnerable \cite{Li2022}. Therefore, to improve a model’s robustness to adversarial data without deteriorating its generalization to regular network traffic flows, it is essential to ensure that adversarial training is performed with realistic examples that could be transmitted through a communication network and preserve the functionality and malicious purpose of a cyber-attack \cite{Rosenberg2021,Vitorino2023}.


\section{Future Directions}
\label{sec:future}

Adversarial data perturbations are very concerning for ML security and reliability. Despite the current difficulty in the transferability of the patch-like and mask-like crafting processes of the image classification domain to the NID domain, there are more sophisticated approaches being specifically designed for communication networks. Nonetheless, an adversarial example that successfully deceives an ML model is not guaranteed to be a successful cyber-attack in a real communication network.

By inspecting the third example of Figure \ref{fig:advflow}, it can be observed that the reason it is impossible is because it does not comply with the inherent data structure of a network traffic flow. On the other hand, the reason that the second example is harmless is because it does not comply with the intended functionality of the Slowloris class of DoS attacks. Therefore, it is possible to define two fundamental properties that are required for an adversarial example to resemble a real data sample:

\begin{itemize}
\item Validity: Compliance with the constraints of a domain, following its inherent data structure.
\item Coherence: Compliance with the constraints of a specific class, following the characteristics that distinguish it from other classes.
\end{itemize}

Even though validity was already taken into account in the reviewed adversarial methods, it is imperative to address it together with coherence to fully achieve adversarial realism. Hence, to ensure that their experiments are realistic, researchers must use examples that represent valid network traffic capable of being transmitted through a real communication network, as well as coherent cyber-attack flows capable of fulfilling their intended functionality and the malicious purposes of an attacker. Novel approaches should support more rigorous configurations to address the complex constraints of the tabular data format and of the time-related characteristics of network traffic flows, creating constrained adversarial examples capable of evading detection while preserving their realism.

As more realistic perturbation crafting processes are developed, they may be used by data scientists and engineers to improve their AI applications with high-quality data, but they may also be used by attackers with malicious intents to disrupt the critical business processes of an organization. An attacker will usually have limited knowledge of the model and feature set utilized in a NID system, corresponding to a black-box or gray-box setting, and will only be able to interact with it in a decision-based approach, without direct feedback and only knowing if a certain example deceived a model if the entire cyber-attack is successfully completed. This limits the possible attack vectors and hinders the feasibility of most adversarial methods of the current literature, but novel attack methods may be developed to address the constraints of complex communication networks and deceive the ML models of NID systems in these scenarios.

To tackle the growing ML attack surface and counteract the disruptions caused by the known attacks, it is becoming essential to enforce a security-by-design approach throughout the entire lifecycle of intelligent systems by simulating realistic evasion attack vectors to assess a system's resilience in edge cases. Reactive and proactive defenses should be combined to attempt to encapsulate ML models in secure software layers and also improve their intrinsic robustness against faulty input data.

Still, efforts to improve a model's robustness against all possible adversarial examples that might occur should not disregard the importance of the model's generalization to regular network traffic that it will definitely encounter. It may not be possible to fully safeguard ML models from adversarial cyber-attack examples, but ML engineers and security practitioners should stay up-to-update with security best-practices and preemptively test their ML models against the novel threats they may face.


\section{Conclusion}
\label{sec:conclsn}

This work investigated the adversarial ML approaches of the current scientific literature that use realistic adversarial examples and could be applied in real ML development and deployment scenarios in the NID domain, fulfilling the requirements of the diverse cyber-attack classes and of the tabular data format and time-related characteristics of network traffic flows.

From the 936 records initially retrieved, over 75\% did not present novel methods nor strategies, which demonstrates that it is difficult for researchers to find innovative approaches in a regular search. The 82 records that remained after the screening phase, which correspond to approximately 9\% of the found publications, presented relevant advances in the use of adversarial ML in the NID domain, and were included in the review together with 16 additional records found through backward snowballing.

The information present in the 98 publications was consolidated and combined into three sections, each discussing its respective RQ, systematizing the approaches, and highlighting the most relevant aspects. An additional section described the open challenges regarding the main RQ, defined the fundamental properties that are required for an adversarial example to be realistic, and provided guidelines for researchers to ensure that their future experiments are adequate for a real communication network.

A security-by-design approach throughout the entire ML lifecycle is becoming essential to tackle the growing ML attack surface, and research efforts continue to be made to better protect various types of algorithms with reactive and proactive defenses. However, there is still a lack of realism in the state-of-the-art perturbation crafting processes, which hinders the development of secure defense strategies. It is pertinent to continue the research efforts to improve the robustness and trustworthiness of ML and of the intelligent cybersecurity solutions that rely on it.

\paragraph{Author Contributions}

Conceptualization, J.V.; methodology, J.V. and E.M.; validation, E.M. and I.P.; investigation, J.V., E.M. and I.P.; writing, J.V.; supervision, I.P.; project administration, I.P.; funding acquisition, I.P. All authors have read and agreed to the published version of the manuscript.

\paragraph{Funding}

The present work was partially supported by the Norte Portugal Regional Operational Programme (NORTE 2020), under the PORTUGAL 2020 Partnership Agreement, through the European Regional Development Fund (ERDF), within project "Cybers SeC IP" (NORTE-01-0145-FEDER-000044). This work has also received funding from UIDB/00760/2020.

\paragraph{Conflicts of Interest}

The authors declare no conflict of interest.

\bibliographystyle{elsarticle-num} 
\bibliography{thebibliography}


\end{document}